\documentclass[a4paper,11pt]{article}
\usepackage{jcappub} 
\usepackage{lineno}

\usepackage[T1]{fontenc}

\usepackage{graphicx}	
\usepackage{amsmath}	

\newcommand{\SISSA}{%
	Theoretical and Scientific Data Science, Scuola Internazionale Superiore di Studi Avanzati (SISSA), 	via Bonomea 265, 34136 Trieste, Italy}

\newcommand{\bbmu}{{\boldsymbol{\mu}}}
\newcommand{\bbtheta}{{\boldsymbol{\theta}}}
\newcommand{\bX}{{\boldsymbol{X}}}
\newcommand{\bx}{{\boldsymbol{x}}}

\newcommand{\bY}{{\boldsymbol{Y}}}
\newcommand{\bb}{{\boldsymbol{b}}}
\newcommand{\bt}{{\boldsymbol{t}}}
\newcommand{\bF}{{\boldsymbol{F}}}
\newcommand{\bJ}{{\boldsymbol{J}}}
\newcommand{\bnabla}{{\boldsymbol{\nabla}}}
\newcommand{\Cm}{{\sf{C}}}
\newcommand{\B}{{\sf{B}}}
\newcommand{\sfA}{{\sf{A}}}

\newcommand{\lnL}{{\mathcal{L}}}

\title{\boldmath Extreme data compression for Bayesian model comparison}

\author[a]{Alan F. Heavens,\note{Corresponding author.}}
\author[b]{Arrykrishna Mootoovaloo,}
\author[a,c,d]{Roberto Trotta,}
\author[e,f]{Elena Sellentin}

\affiliation[a]{Imperial Centre for Inference and Cosmology (ICIC), Department of Physics, Imperial College, Blackett Laboratory, Prince Consort Road, London SW7 2AZ, U.K.}
\affiliation[b]{Department of Physics, University of Oxford, Denys Wilkinson Building, Keble Road, Oxford OX1 3RH,
U.K.}
\affiliation[c]{\SISSA}
\affiliation[d]{Centro Nazionale ``High Performance Computer, Big Data and Quantum Computing''}
\affiliation[e]{Mathematical Institute, Leiden University, Snellius Gebouw, Niels Bohrweg 1, NL-2333 CA Leiden, The Netherlands}
\affiliation[f]{Leiden Observatory, Leiden University, Oort Gebouw, Niels Bohrweg 2, NL-2333 CA Leiden, The Netherlands}

\emailAdd{a.heavens@imperial.ac.uk}

\abstract{
We develop extreme data compression for use in Bayesian model comparison via the MOPED algorithm, as well as more general score compression.  We find that Bayes factors from data compressed with the MOPED algorithm are identical to those from their uncompressed datasets when the models are linear and the errors Gaussian. In other nonlinear cases, whether nested or not, we find negligible differences in the Bayes factors, and show this explicitly for the Pantheon-SH0ES supernova dataset.  We also investigate the sampling properties of the Bayesian Evidence as a frequentist statistic, and find that extreme data compression reduces the sampling variance of the Evidence, but has no impact on the sampling distribution of Bayes factors.  Since model comparison can be a very computationally-intensive task, MOPED extreme data compression may present significant advantages in computational time.}

\begin{document}
\maketitle
\flushbottom



\section{Introduction}

Astronomical datasets can be very large, with galaxy catalogues and weak lensing surveys having up to hundreds of millions of objects \citep[e.g.,][]{KiDS2019,DES_shear_2022,HSC2023}, and Planck microwave background data \cite{Planck_2020_1} containing billions of time-ordered data.  Extracting scientific knowledge from such large datasets almost inevitably requires some form of data compression, typically into summary statistics such as point estimates of two-point summaries, either correlation functions or power spectra.   Compression without loss of information is generally possible since the datasets are noisy and correlated, and furthermore the information may be captured by a small number of summary statistics, depending on the statistical properties of the underlying data.  However, even with compression to two-point summaries, the number of statistics can still be large: for example, cosmic shear surveys such as proposed for the Euclid mission \citep{Laureijs} may generate $\sim 10^4$ two-point statistics, and this can present problems for likelihood-based scientific analysis if the covariance of the summaries needs to be simulated, or indeed if their distributions are not known \citep{Heavens2017}.  Both of these considerations motivate the use of more radical data compression techniques. This can also be highly beneficial when the resultant summaries are coupled with techniques such as simulation-based inference (SBI, also known as likelihood-free inference), which is undergoing rapid development thanks in part to novel machine learning techniques \citep[e.g.,][]{IMNN,Gerardi2021,Makinen2021,Jeffrey2021,Makinen2022}.  Typically, the most extreme compression possible without losing information is to reduce the dataset down from its original size to the number of model parameters to be inferred.  Such extreme compression can have a number of benefits, in terms of speed of analysis, in reducing the number of simulations required to determine the sampling distribution of the statistics \citep{Heavens2017}, and in making SBI feasible \citep{Lin2022, Leclercq2022}, whilst still allowing for new models to be considered \citep{Heavens2020}.  Heavens et al. \cite{2000MNRAS.317..965H} derived the optimal data compression algorithm MOPED, for gaussian-distributed data whose mean signal depends on the parameters, showing that extreme data compression can preserve the Fisher matrix, so the compression is, in this sense, lossless.  Techniques, based on Karh\" unen-Lo\`eve compression \citep{TTH} also exist for data whose covariance depends on the parameters, but not the mean, in contrast to MOPED, which applies when the mean depends on the parameters, but not the covariance. However, the former situation does not lead to lossless extreme compression, nor does it generalise neatly to multiple parameters. MOPED, whether in its original Gram-Schmidt orthogonal form, or in the simpler non-orthogonalised version (which we use in this paper) is a special case of using the score function (the gradient of the likelihood with respect to the model parameters) to compress the data, as shown by \cite{AlsingWandelt2018}.  Alternative approaches to extreme data compression include information-maximising neural networks (IMNNs) to find maximally-informative summaries, as introduced by \cite{IMNN}, and which have in certain cases been shown to be lossless at Fisher level \citep{Makinen2021}.  Typically, the neural network-based methods operate on the fields themselves \citep[e.g.][]{Jeffrey2021,Makinen2021,Makinen2022}, whereas the MOPED and score compressions operate on some intermediate, physics-based summaries \cite[e.g.][]{Heavens2017,Jeffrey2021,Jung2022,Lin2022}.

The main motivation for this paper is to determine whether the qualities of optimal extreme data compression which apply to parameter inference also extend to Bayesian Evidence and Bayes factors, i.e., to model comparison. The answer to this question is yes, provided that the extreme data compression is performed in each of the models separately, and the compressed data then combined, such that the combined compressed dataset is by construction optimal in all models. This applies to both nested and non-nested models. 

With this key result, committed Bayesians may stop reading, but we also take the opportunity to investigate the sampling distribution of the Bayesian Evidence and the related Bayes Factor, to look at their use as frequentist statistics under extreme compression.  The motivation for this comes from a suggestion \citep{Joachimi21} that data compression may help in reducing the sampling variability in the Bayesian Evidence.  The answer is that it does, but the sampling variance of the Bayes Factor, which is a ratio of evidences and the relevant quantity for model comparison, remains virtually unchanged.  To rephrase this same point, one can use extreme MOPED data compression as effectively as the full dataset for model comparison as well as parameter inference.  Since model comparison is typically a computationally expensive exercise, extreme data compression, which often results in faster likelihood evaluation, can be very valuable.

Note that the distinction between the sampling distribution of the Bayesian Evidence and of the Bayes Factor is important, especially since it is tempting to conclude --erroneously-- that the sampling distribution of the Bayesian Evidence is sufficiently wide to compromise its usefulness in model selection, but this ignores the fact that much of the variation is common to both models (the data are the same), and cancels out when the ratio of Bayesian Evidence is computed, leaving the Bayes factor as an effective discriminant, as we show in this paper.

The layout of the paper is as follows.  In section \ref{sec:MOPED} we review the MOPED data compression algorithm, and show how it works effectively for model comparison as well as parameter inference, and consider more general score compression.   In section \ref{sec:frequentist} we explore the frequentist properties of Bayes factors, showing that the sampling distribution of the Bayes factor is unaltered whether the full or compressed data are used.  In section \ref{Pantheon} we apply MOPED compression to the Pantheon+SH0ES dataset \citep{Scolnic2022}) and show that both the cosmological parameters and the Bayes Factor can be determined from just 3 numbers (the MOPED coefficients), as precisely as from the original 1590 supernovae.  We conclude in section \ref{sec:conclusions}.

\section{Extreme data compression: Bayesian Evidence with MOPED}
\label{sec:MOPED}

\subsection{Bayesian Evidence and Bayes Factors for the uncompressed dataset}

Let us first consider gaussian-distributed data, with fixed covariance matrix and arbitrary dependence of the mean on the parameters.  Data compression with MOPED \cite{2000MNRAS.317..965H} allows extreme compression to $p$ summary statistics, where $p$ is the number of model parameters, whatever the length of the original data vector.  Although the resulting  MOPED coefficients are not strictly sufficient statistics except in the case of linear models, they can preserve the entire Fisher matrix, so the likelihood surface near the peak is no broader than when using the full, uncompressed dataset.  As a result, the extreme compression can be used for parameter inference without increasing errors, and has been used in many contexts, such as star formation histories of SDSS galaxies \citep[e.g.][]{Heavens2004,Panter2003,Panter2007,Panter2008}, the cosmic microwave background \citep[e.g.][]{Gupta2002,Zablocki2016,Prince2019}, galaxy power spectrum and bispectrum \citep[e.g.][]{Lai2023,Gualdi2019}, gravitational waves \citep{Graff2011}, weak lensing \citep[e.g.][]{Lin2022,Thiele2023}, combined probes \citep[e.g.][]{Ruggeri2020,Wenzl2022}, planetary transits \citep{Protopapas2005}, and extended to parameter-dependent covariance matrices \citep{Heavens2017} and mis-specified models \citep{Heavens2020}. In this section, we explore how MOPED and related score compression can be used in a new context: model comparison.  To begin we review the derivation of the unorthogonalised version of MOPED, following not the original analysis \citep{2000MNRAS.317..965H}, but rather the derivation in \cite{AlsingWandelt2018}.

We consider a model with parameters  $\bbtheta=\{\theta_1, \dots, \theta_p\}$, data $\bx \in \mathbb{R}^n$ (where $n \gg p$) generated from a gaussian sampling distribution with covariance matrix $\Cm \in \mathbb{R}^{n\times n}$ independent of $\bbtheta$ and expectation value of the data $\bbmu(\bbtheta)$. The sampling distribution for the uncompressed dataset is therefore
\begin{equation}
p(\bx | \bbtheta) =\frac{1}{\sqrt{| 2\pi\Cm |}}\exp\left[-\frac{1}{2}(\bx-\bbmu)^T\Cm^{-1}(\bx-\bbmu)\right].
\end{equation}
Letting $\bX \equiv \bx-\bbmu_*$, where $\bbmu_*$ is the expectation value of the data at some fiducial set of parameters $\bbtheta_*$, the posterior is
\begin{equation}
p(\bbtheta|\bx) \propto  \frac{1}{\sqrt{| 2\pi\Cm |}}\exp\left[-\frac{1}{2}(\bX-\tilde\bbmu)^T\Cm^{-1}(\bX-\tilde\bbmu)\right]\, \pi(\bbtheta) 
\end{equation}
for a prior density $\pi(\bbtheta)$.  Here 
\begin{equation}
\tilde\bbmu(\bbtheta) \equiv \bbmu(\bbtheta)-\bbmu_*,
\end{equation}
and a linear Taylor expansion around the fiducial point gives 
\begin{equation}
\tilde\bbmu \simeq \bbmu_{,\alpha} \tilde\bbtheta_\alpha\quad {\rm where}\quad \tilde\bbtheta \equiv \bbtheta-\bbtheta_*.
\label{Taylor}
\end{equation}
The comma indicates $\partial/\partial \bbtheta_\alpha$ and the summation convention is used. Hence to this order,
\begin{align}
p(\bbtheta|\bx) \propto & \frac{1}{\sqrt{| 2\pi\Cm |}}\exp\left(-\frac{1}{2}\bX^T\Cm^{-1}\bX\right) 
\exp\left(\tilde\bbmu^T\Cm^{-1}\bX -\frac{1}{2}\tilde\bbmu^T\Cm^{-1}\tilde\bbmu\right)\, \pi(\bbtheta).
\end{align}
We see that the only coupling of the data $\bx$ to the parameters is through the $\tilde\bbmu^T\Cm^{-1}\bX$ term, and only via the $p$ MOPED coefficients
\begin{equation}
 Y_\alpha \equiv \bbmu_{,\alpha}^T \Cm^{-1} \bX\equiv \bb_\alpha^T (\bx-\bbmu_*),\qquad \alpha=1\ldots p.
\label{eq:MOPEDcomp} 
\end{equation}
and we identify the $p$ (unorthogonalised) MOPED weight vectors  \citep{2000MNRAS.317..965H})
\begin{equation}
\bb_\alpha^T \equiv \bbmu_{,\alpha}^T \Cm^{-1}.
\label{eq:MOPEDvec}
\end{equation}
The posterior may then be written
\begin{align}
p(\bbtheta|\bx) &\propto \frac{1}{\sqrt{| 2\pi\Cm |}}\exp\left(-\frac{1}{2}\bX^T\Cm^{-1}\bX\right) 
\exp\left[(\bb_\alpha^T \bX)\tilde\theta_\alpha -\frac{1}{2} (\bb_\alpha^T\Cm \bb_\beta) \tilde\theta_\alpha \tilde\theta_\beta\right]\, \pi(\tilde\bbtheta),
\label{eq:posterior}
\end{align}
or in matrix form
\begin{align}
p(\bbtheta|\bx) &\propto \frac{1}{\sqrt{| 2\pi\Cm |}}\exp\left(-\frac{1}{2}\bX^T\Cm^{-1}\bX\right)  
\exp\left(\bY^T\tilde\bbtheta -\frac{1}{2}\tilde\bbtheta^T\Lambda \tilde\bbtheta \right) \pi(\tilde\bbtheta),
\label{eq:posteriorMatrix}
\end{align}
where $\bY$ is a vector of only $p$ elements, consisting of the MOPED coefficients:
\begin{equation}
    \bY = \B^T \bX = \B^T (\bx-\bbmu_*),
\end{equation}
and $\B^T \in \mathbb{R}^{p\times n}$ is made up of $p$ rows of MOPED vectors $\bb_\alpha^T$, defined by equation (\ref{eq:MOPEDvec}).  In matrix form,
\begin{equation}
    \B^T = \Phi^T \Cm^{-1},
\end{equation}
where $\Phi^T \in \mathbb{R}^{p\times n}$ is a matrix with rows given by $\bbmu_{,\alpha}^T$.  
The symmetric matrix $\Lambda \in \mathbb{R}^{p \times p}$ is
\begin{equation}
    \Lambda = \B^T\Cm\B = \Phi^T \Cm^{-1} \Phi.
\end{equation}
Note that Heavens et al. \cite{2000MNRAS.317..965H} applied a Gram-Schmidt orthogonalisation to the MOPED vectors, to decorrelate the elements of $\bY$, but the information content is the same if this is not applied, and we omit it as it simplifies the analysis.  We have also subtracted $\bbmu_*$ from $\bx$ for convenience, and to ensure that a Taylor expansion to linear order is accurate, following \cite{Heavens2017}.   

To normalise the posterior, $p(\bbtheta | \bx,M) = p(\bx|\bbtheta,M)\pi(\bbtheta)/p(\bx|M)$, we need the Bayesian Evidence $Z(\bx) = p(\bx|M)$, where we have added explicitly the dependence on the choice of model, $M$.  The Evidence is
\begin{align}
p(\bx|M) &=\frac{1}{\sqrt{| 2\pi\Cm |}}\exp\left(-\frac{1}{2}\bX^T\Cm^{-1}\bX \right)  
\int\,\exp\left(\bY^T\tilde\bbtheta -\frac{1}{2}\tilde\bbtheta^T\Lambda \tilde\bbtheta \right) \pi(\tilde\bbtheta)\,d\tilde\bbtheta.
\label{eq:ZX}
\end{align}
We see that the data-only prefactor cancels in the normalized posterior:
\begin{align}
p(\bbtheta|\bx,M) &= \frac{
\exp\left(\bY^T\tilde\bbtheta -\frac{1}{2}\tilde\bbtheta^T\Lambda \tilde\bbtheta \right) \pi(\tilde\bbtheta|M)}{\int\,\exp\left(\bY^T\tilde\bbtheta -\frac{1}{2}\tilde\bbtheta^T\Lambda \tilde\bbtheta \right) \pi(\tilde\bbtheta|M)\,d\tilde\bbtheta}.
\end{align}
From this it is apparent that, to the extent that the linear expansion is applicable, we may compute the posterior from the MOPED coefficients alone, without considering $\bx$ in full --- we only need ${\bf Y}$.  This is a very powerful feature of MOPED compression.

Note that for a Gaussian prior density, $\pi(\tilde\bbtheta|M) =  {\cal {N}}({\bf 0},\Sigma)$ the integrals may be evaluated analytically (see also \cite{2008ConPh..49...71T,Leclercq2019,Leclercq2022}) giving
\begin{align}
Z(\bX) &=\frac{1}{\sqrt{| 2\pi\Cm ||\Sigma\Lambda+\mathbb{I}_p|}}\exp\left(-\frac{1}{2}\bX^T\Cm^{-1}\bX\right) 
\exp\left[\frac{1}{2}\bY^T\left(\Lambda+\Sigma^{-1}\right)^{-1} \bY\right].
\label{eq:ZXG}
\end{align}

\subsubsection{Demonstration that the Bayes factor depends only on MOPED coefficients}

Before we derive the general expression for the Bayes factor in the next section, we motivate our approach by demonstrating that, for
nested models, the Bayes factor from the full
dataset depends only on the set of MOPED coefficients defined in the extended model, to linear order in expansion. The posterior odds between models $M_0$ and $M_1$ are
\begin{equation}
    \frac{p(M_0 | \bx)}{p(M_1 | \bx)} = \frac{\pi(M_0)p(\bx|M_0)}{\pi(M_1)p(\bx|M_1)} = B_{01}(\bx) \equiv \frac{Z_0(\bx)}{Z_1(\bx)}
\end{equation}
where $B_{01}$ is the Bayes factor, 
and the last equality applies if we assume equal prior model probabilities, $\pi(M_0)=\pi(M_1)$. We have defined $Z_i \equiv p(\bx|M_i)$.

Model $M_0$, with $q$ free parameters, is said to be `nested' in model $M_1$, with $p>q$ parameters, if the first $q$ parameters of Model 1 are common to Model 0, and there is a choice of value for the additional parameters in Model 1 such that the latter reverts to Model 0. In this setting, we choose a fiducial parameter set within the prior support of Model 0 (and, therefore, {\it a fortiori} in Model 1); in this case, $\bbmu_*$ is common to both models, and $\bX$ is the same in both. in this case, the prefactor in the Evidence (equation \ref{eq:ZX}), cancels out in the Bayes factor, for any choice of prior densities $\pi(\tilde\bbtheta_i|M_i)$
\begin{equation}
    B_{01}(\bx) \equiv \frac{Z_0(\bx)}{Z_1(\bx)} = 
    \frac{\int\,\exp\left(\bY^T\tilde\bbtheta_0 -\frac{1}{2}\tilde\bbtheta_0^T\Lambda_0 \tilde\bbtheta_0 \right) \pi(\tilde\bbtheta_0|M_0)\,d\tilde\bbtheta_0}{\int\,\exp\left(\bY^T\tilde\bbtheta_1 -\frac{1}{2}\tilde\bbtheta_1^T\Lambda_1 \tilde\bbtheta_1 \right) \pi(\tilde\bbtheta_1|M_1)\,d\tilde\bbtheta_1}.
    \label{eq:B01x}
\end{equation}
Note that for the
simpler model the extra parameters are set at $\tilde\bbtheta={\bf 0}$, which we accommodate by defining $\Lambda_0$ as the top $q\times q$ block of $\Lambda_1$. 

Just like the posterior for the parameters, the Bayes factor for the full dataset depends on the data only via $\bY$.  This motivates adopting MOPED compression for model comparison in the general case, with the potential for a large acceleration of what can be a computationally expensive task for some applications.  We develop this in the next section.  

It is worth pointing out that the requirements on the accuracy of the Taylor expansion are more stringent for model comparison than for parameter inference, since we are evaluating an integral over the whole parameter space, well beyond the region where the posterior density is appreciable.  For some applications this could limit the accuracy of the approach, but for typical cosmological posteriors, their unimodal and relatively simple structure means that the MOPED-derived posteriors are often very close to those obtained from the full dataset, even far from the posterior peak.  This is apparent even in the first (astrophysical) example, presented in \cite{2000MNRAS.317..965H}, where the posterior contours are similar between MOPED and full data set at values smaller than the peak by a factor $e^{100}$, even though this is not guaranteed by the method - the compression does far better than one has any right to expect. We therefore compress the data only once for each model, each at a single fiducial parameter set, and we illustrate later in a cosmological supernova study that the Bayes factor computation from the compressed data still agrees with the full dataset within sampling errors.

\subsection{Bayesian Evidence and Bayes Factor with MOPED compression for non-nested models}
\label{sec:MOPEDZ}

We now develop the formalism for using MOPED extreme data compression for computing Bayes factors for non-nested models, with the nested case requiring only a minimal adjustment. In order for the Bayes factor to give the correct Bayesian posterior odds ratio (for equal model priors), the data entering in each of the Bayesian Evidence calculations have to be the same, or else the $p({\rm data})$ denominator in Bayes theorem will not cancel in the ratio.   Since the MOPED vectors defined within each model are optimal for parameter inference in their respective models, we construct a compressed dataset from the union of the MOPED coefficients from both (or, more generally, all) models.  In this way the augmented compressed dataset will be optimal for parameter inference in both models, since we will not lose information by adding additional data beyond what is required. Of course we must take care to include correctly the correlations between all the compressed data.   We thus compare the analysis with the full dataset with the data compressed to 
\begin{equation}
    \bY^T = 
    \begin{pmatrix}
        \bY_0^T\\
        \bY_1^T\\
    \end{pmatrix}
    =
    \begin{pmatrix}
            \B_0^T(\bx-\bbmu_{*0})\\
            \B_1^T(\bx-\bbmu_{*1})\\
         \end{pmatrix} \in \mathbb{R}^{p+q},
\end{equation}
where $\bbmu_{*0}$ and $\bbmu_{*1}$ are fiducial means in the two models.  In the nested case, these would be chosen to be equal, and since $\bY_0$ is already contained in $\bY_1$, it is discarded, with $\bY_1$ alone being retained. Here we assume non-nested models when both $\bY_0$ and $\bY_1$ are present, and specialize to the nested case later. 

The Bayesian Evidence for each model, $Z_i({\bY})$ is obtained by very similar arguments to the preceding analysis.  For Gaussian data $\bx \sim {\cal N}(\bbmu, \Cm)$, $\bY \sim {\cal N}(\bar\bY, \Lambda) $ is also Gaussian-distributed, with mean
\begin{equation}
    \bar\bY(\tilde\bbtheta) =     
        \begin{pmatrix}
        \B_0^T(\bbmu-\bbmu_{*0})\\
        \B_1^T(\bbmu-\bbmu_{*1})\\
    \end{pmatrix}.
\end{equation}
and covariance $\Lambda = \Phi^T \Cm^{-1} \Phi \in \mathbb{R}^{(p+q) \times (p+q)}$, where in block form $\Phi = (\Phi_0,\Phi_1)$.  The Bayesian Evidence for model $i$ is then
\begin{equation}
    Z_i(\bY) = \frac{1}{\sqrt{|2\pi \Lambda|}} 
    \int \exp\left[-\frac{1}{2}
    \left(\bY-\bar\bY\right)^T \Lambda^{-1}
    \left(\bY-\bar\bY\right)
    \right] \, \pi(\tilde\bbtheta_i|M_i)\, d\tilde\bbtheta_i.
\end{equation}
With a Taylor expansion for $\bbmu$ as before (equation (\ref{Taylor})), $\bar\bY = \B^T \Phi \tilde\bbtheta \equiv \Lambda \tilde\bbtheta$, we find
\begin{align}
    Z_i(\bY)& =
    \frac{1}{\sqrt{|2\pi\Lambda|}}
    \exp\left[-\frac{1}{2}(\bY-\bar\bY_{*i})^T\Lambda^{-1}(\bY-\bar\bY_{*i})\right] \times \nonumber\\
    &\int \exp\left[(\bY-\bar\bY_{*i})^T \Lambda^{-1} \sfA_i \tilde\bbtheta_i 
    -\frac{1}{2}\tilde\bbtheta_i^T \sfA_i ^T \Lambda^{-1}  \sfA_i \tilde\bbtheta_i\right]  \pi(\tilde\bbtheta_i|M_i) d\tilde\bbtheta_i,
    \label{ZMOPED}
\end{align}
where $\sfA_i \equiv \Phi^T \Cm^{-1}\Phi_i$ and 
\begin{equation}
    \bar\bY_{*i} = \begin{pmatrix}
            \B_0^T(\bbmu_{*i}-\bbmu_{*0})\\
            \B_1^T(\bbmu_{*i}-\bbmu_{*1})\\
         \end{pmatrix},
\end{equation}
which will have zeros in one or other block depending on whether $i=0$ or $1$.

\subsubsection{Nested models with MOPED compression}

There is a significant simplification for nested models, where we dispense with the top blocks of $\bY$ and $\Phi^T$, and take $\bbmu_{*0}=\bbmu_{*1}$. As a consequence $\bar\bY_{*i}={\bf 0}$.  $\sfA_i$ can be replaced by $\Lambda$ (which reduces to $\Lambda_1)$, since it always comes in combination with $\tilde\bbtheta_i$, and we extend the $\tilde\bbtheta_0$ vector to include zeros from the extended parameter space.  With these simplifications, the Evidence becomes
\begin{align}
    Z_i(\bY)& =
    \frac{1}{\sqrt{|2\pi\Lambda|}}
    \exp\left(-\frac{1}{2}\bY^T\Lambda^{-1}\bY\right) 
    \int \exp\left(\bY^T  \tilde\bbtheta_i - \frac{1}{2}\tilde\bbtheta_i^T \Lambda  \tilde\bbtheta_i\right)  \pi(\tilde\bbtheta_i|M_i) d\tilde\bbtheta_i.
    \label{ZMOPED_nested}
\end{align}
This is very similar to the Evidence computed from the full data set (equation \ref{eq:ZX}), but with a different data-dependent prefactor to the integral. 

When we set the prior on the extended model parameters to Dirac delta functions in Model 0, $\Lambda$ effectively becomes $\Lambda_0$, and when we cancel the common prefactors, we recover equation (\ref{eq:B01x}), and find that the Bayes factor for the MOPED compressed data for nested models is exactly the same as the Bayes factor for the original dataset.  In other words, to the extent that a linear Taylor expansion is accurate, no information is lost in performing Bayesian model comparison using MOPED coefficients rather than the full dataset.

\subsubsection{Gaussian prior}
\label{sec:gaussiannonnested}
With a gaussian prior on the parameters, the Bayesian Evidence (equation \ref{ZMOPED}) for the different models (nested or otherwise) reduces, after application of the Woodbury formula and the matrix determinant lemma, to a very simple expression 
\begin{equation}
    Z_i({\bY}) = \frac{1}{\sqrt{|2\pi Q_i|}}\exp\left[ -\frac{1}{2}(\bY-\bar\bY_{*i})^T Q_i^{-1}(\bY-\bar\bY_{*i})\right]
\end{equation}
where
\begin{equation}
Q_i = \Lambda + \sfA_i^T \Sigma_i \sfA_i.
\end{equation}

\subsection{Score compression}
\label{sec:score}

We can generalise the result further to arbitrary likelihood functions, using the score compression of Alsing \& Wandelt \cite{AlsingWandelt2018}, of which MOPED is a special case. Here they expand the log likelihood as 
\begin{equation}
\lnL = \lnL_* + \tilde\bbtheta^T \bnabla \lnL_* - \frac{1}{2}\tilde\bbtheta^T \bJ_* \tilde\bbtheta
\end{equation}
where $\bJ \equiv -\bnabla\bnabla^T\lnL$, and $*$ indicates that quantities are evaluated at the fiducial parameters.

$\lnL_*$ depends on the data, but we see again that the $\lnL_*$ (which is a source of variability in the Bayesian Evidence) cancels in the Bayes factor if the models are nested and $\bbtheta_*$ lies in the domain of both $M_0$ and $M_1$. The only coupling between data and parameters at linear order is through the score function
\begin{equation}
\bt \equiv \bnabla\lnL_*.
\end{equation}
${\bf J}$ is strictly constant in the Gaussian case if $\Cm$ is independent of $\bbtheta$, and is given by the Fisher matrix 
\begin{equation}
\langle \bJ \rangle = \bF = \bnabla\bbmu^T \Cm \bnabla^T\bbmu.
\label{J}
\end{equation}
If $\Cm$ is parameter-dependent, the expression is complicated, and couples data and parameters (see \cite{AlsingWandelt2018}, equation (15)), and replacing $\bJ$ by $\bF$ would be an approximation, but possibly a good one.

We see that the preceding arguments hold in this case as well, that for nested models, the $\lnL_*$ term cancels out in both the posterior and also the Bayes factor, if the fiducial parameters are common to both (all) models.  As a result, general score compression, using the trick of concatenating the scores from all models considered, can effectively be used for model comparison, with the additional approximation (\ref{J}).

\section{Example Applications}

\subsection{Distribution of Bayes factors}
\label{sec:frequentist}

Bayesians may wish to turn away now, as we consider the frequentist properties of Bayesian Evidence and Bayes factors, as investigated by Jenkins \& Peacock  \cite{JP11} and Joachimi et al. \cite{Joachimi21}. These papers note that the Bayesian Evidence and the Bayes factor are noisy statistics, and consider their sampling variability and its impact on the outcome of model comparison.   Note that we remain of the Bayesian view that the Bayes factor can indicate a preference for one model over another, but not falsify models in absolute terms, and we are not seeking frequentist methods to do this, which typically rely on tail probabilities of $p({\rm data}|{M})$ as opposed to posterior model probabilities, $p({M}|{\rm data})$, cf. \cite{RN565,Koo2022,Keeley2022}.

To the committed Bayesian, the sampling distribution is not relevant, since the Bayesian view is that there is one set of data on which the inference is based, and it does not matter what other realisations of the data might have yielded.  Nevertheless, $Z$ is a statistic, and its sampling distribution could be of some interest. There is a contribution to the variability of $Z$ from the data-only prefactor in equation (\ref{eq:ZX}) in the gaussian case (and we come to a similar conclusion in the more general score compression case: see section \ref{sec:score}).  The log of the prefactor is distributed as a
$\chi^2$ distribution with $n$ degrees of freedom with a variance of $2n$.  However, this main source of variability (the prefactor) cancels in nested models when the Bayes factor is computed, so the quantity on which model comparison depends is not subject to this source of large variability.

This behaviour is apparent in Fig. B.1 of \cite{Joachimi21}, where the Bayes factor is very stable even though $Z_0$ and $Z_1$ are highly variable, which would invite the na\"ive interpretation that the sampling variance of the Evidence may hinder model comparison. However, even if the sampling distribution of $\ln Z_0$ and $\ln Z_1$ is large in comparison with $\ln B_{01}(\bX)$, the Bayes factor favours one model consistently, because $Z_0$ and $Z_1$ are not independent, as they use the same data realisation.

\begin{figure}%
    \centering
    \includegraphics[width=0.48\textwidth]{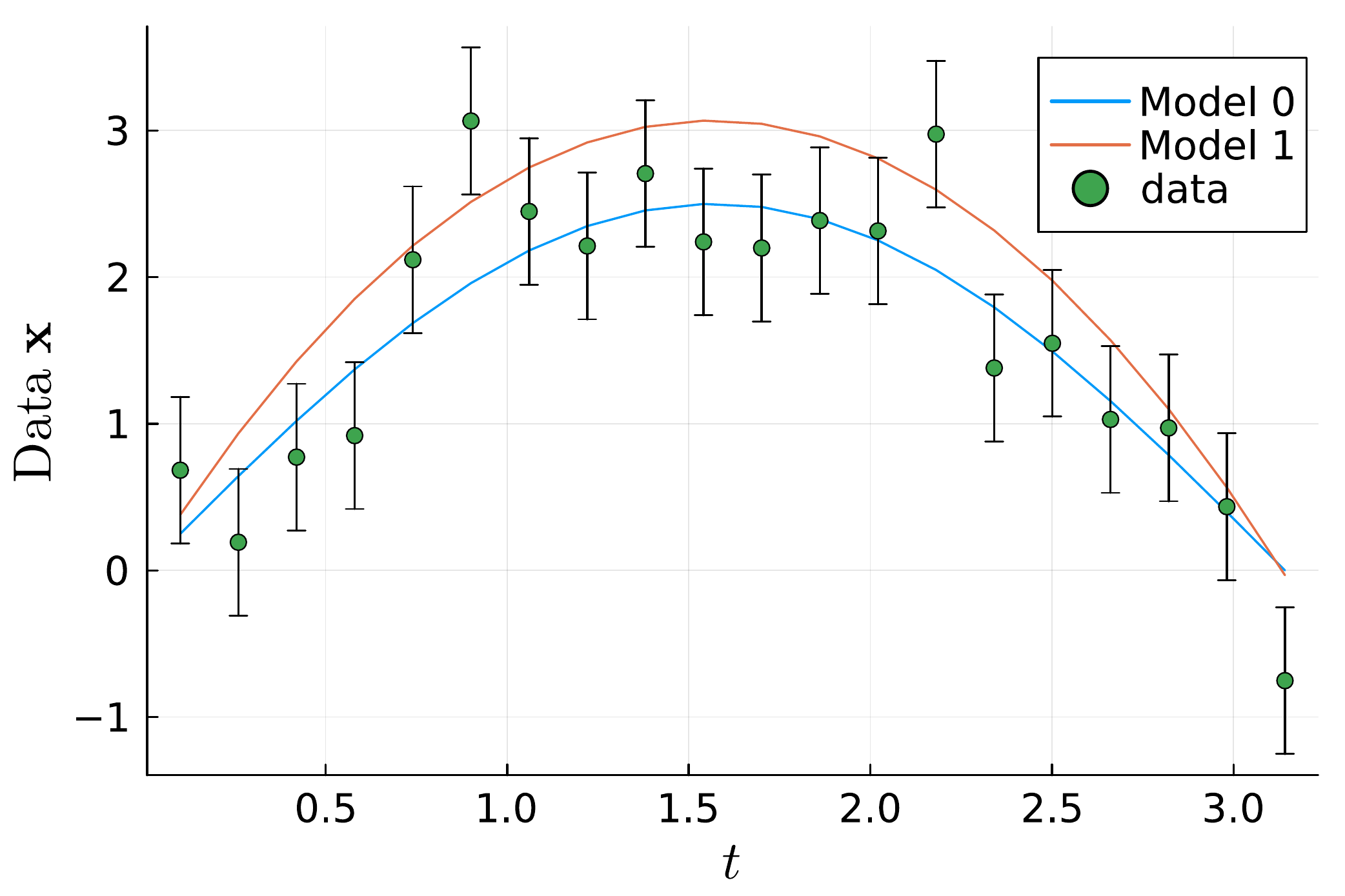}
    \caption{Example dataset, with noisy data drawn from the mean curve of Model 0. Parameters for Model 1 are arbitrary.}
    \label{fig:data}%
\end{figure}

We illustrate the behaviour of a non-nested case with the following toy non-linear models:
\begin{itemize}
\item{$M_0:\ x = E \sin(F t);\quad \bbtheta_0 = (E,F) $}
\item{$M_1:\ x = a t^2 + b t; \quad \bbtheta_1 = (a,b) $}
\end{itemize}
where $t$ has $20$ values uniformly spaced between $0$ and $\pi$ (an example is shown in Fig. \ref{fig:data}).  

\begin{figure}%
    \centering
    \includegraphics[width=0.48\textwidth]{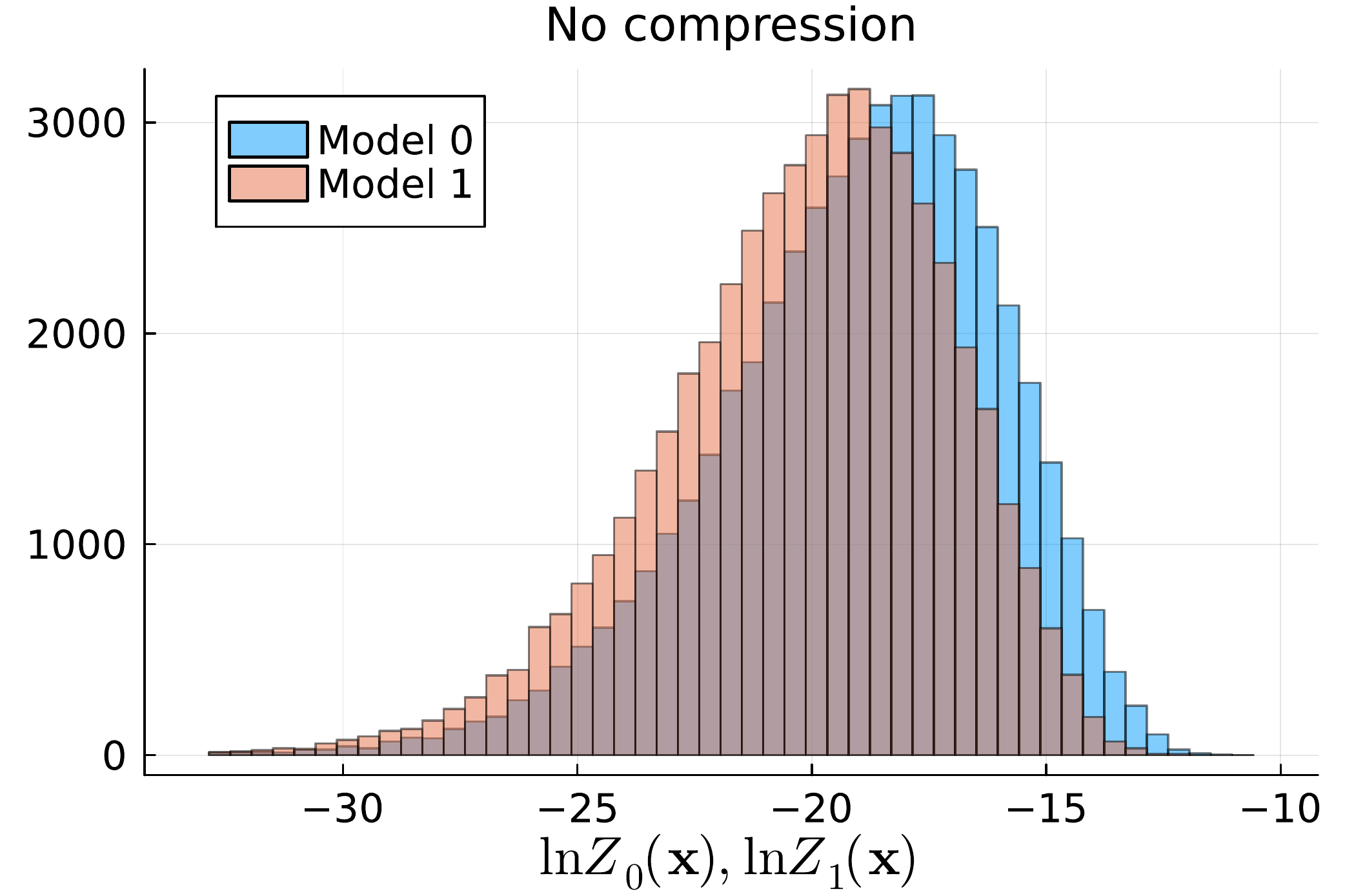}
    \caption{The distribution of Bayesian Evidence for $50000$ noise realisations of non-nested toy models $M_0$ and $M_1$ with details given in the text, obtained from the full dataset.  The normalisation of the $y$ axis is arbitrary.}%
    \label{fig:ZX}%
\end{figure}

\begin{figure}%
    \centering
    \includegraphics[width=0.48\textwidth]{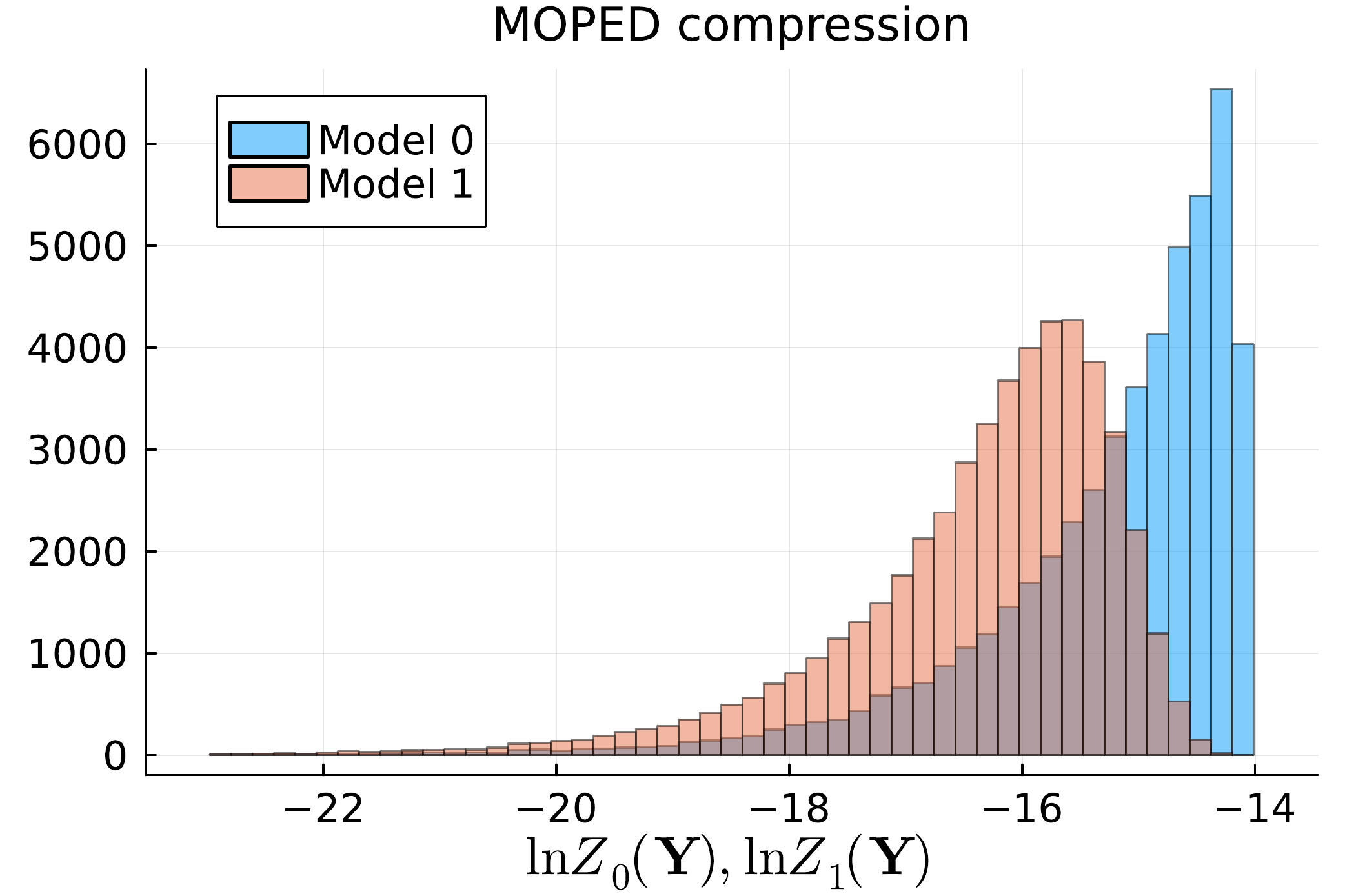}
    \caption{As Fig. \ref{fig:ZX}, but for a compressed dataset consisting of the four MOPED coefficients, two  from each model.}
    \label{fig:ZY}%
\end{figure}

The data are generated with $E=2.5$ and $ F=1.0$, and the fiducial values are $E_* = 3.0$, $F_*=1.2, a_*=-1.5, b_*=4.7$).  The distribution of Bayesian Evidence values for each model from $4 \times 10^4$ realisations is shown in Fig. \ref{fig:ZX} for the full dataset, and from MOPED-compressed data in Fig. \ref{fig:ZY}.  Notice that the considerable sample variance exceeds the difference of the mean Evidence values.  As suggested by \citep{Joachimi21}, the width of the sampling distribution of the compressed data Evidence is reduced compared to the full dataset. 

\begin{figure}%
    \centering
    \includegraphics[width=0.48\textwidth]{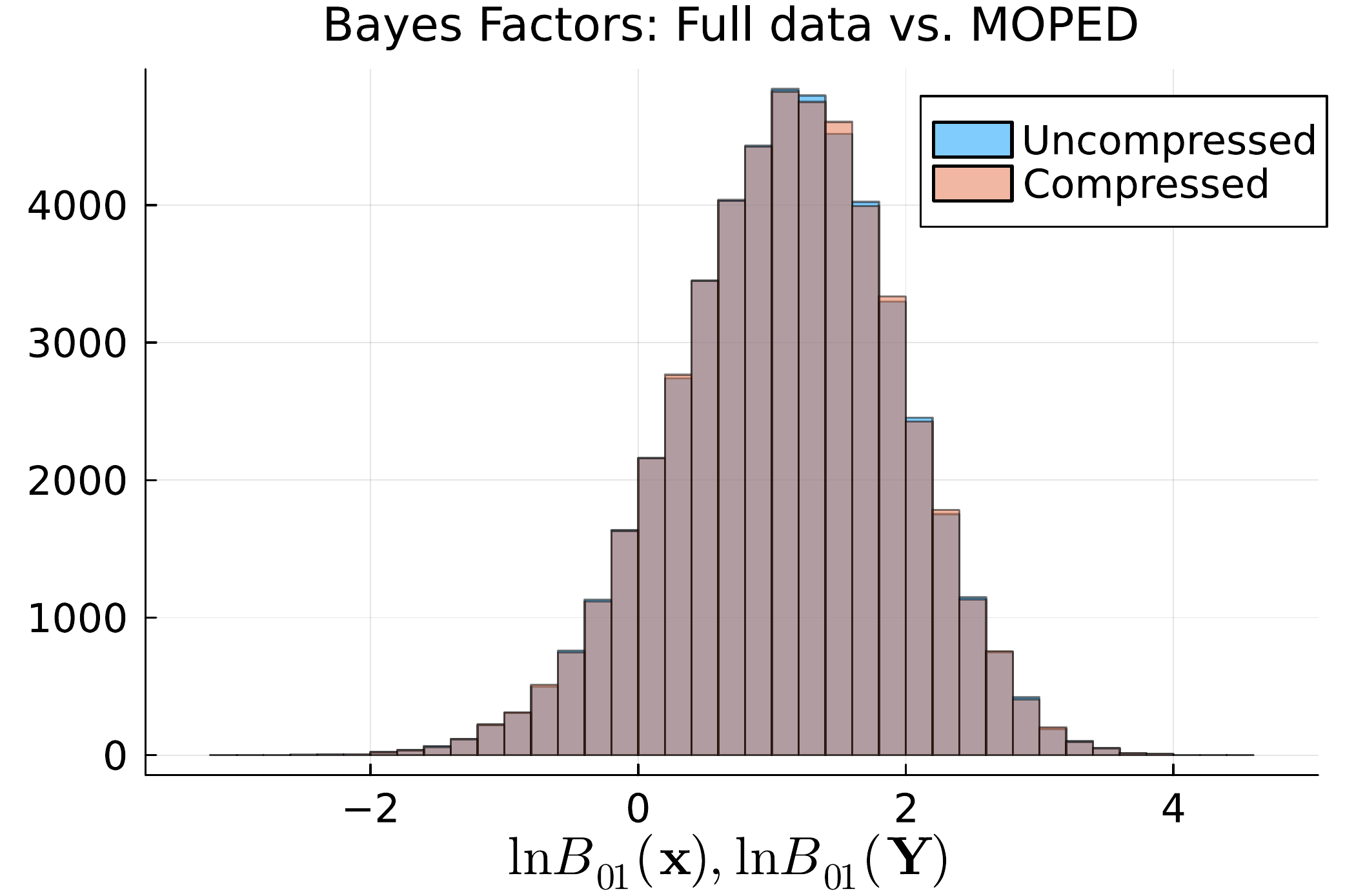}
    \caption{Comparison of Bayes factor sampling distribution for the nonlinear, non-nested toy models $M_0$ and $M_1$ with details given in the text, from the full dataset and the MOPED-compressed dataset. The distributions overlay almost perfectly. }
    \label{fig:BXY_nonlin}
\end{figure}

Fig. \ref{fig:BXY_nonlin} shows that, although the Evidence values differ markedly for the full and compressed datasets, the Bayes factors are almost identical, with deviations being due to the linear Taylor expansion not being perfect in the nonlinear case. 

 Our conclusion from this study is that MOPED compression reduces the variance of the sampling distribution of the Bayesian Evidence, so it would appear advantageous to use such a data compression.  However, the Evidence values from an individual noise realisation are highly correlated, as they are based on the same data, and the log of the Bayes factor has a much smaller variance than the variance of the individual $\ln Z$ values. We find that the uncompressed data give the same Bayes factor as the MOPED compressed data (inasmuch as the neglect of higher-order derivatives of $\bbmu$ is valid).  As a result MOPED compression can be used for accurate model comparison, with in some cases a large increase in speed of computation.   We also note that with a single full likelihood evaluation and one compressed likelihood evaluation at the fiducial parameters, the full dataset Evidence can be computed, if it were ever needed, from the compressed Evidence:
 \[
 Z(\bX) = \frac{\sqrt{|2\pi \Lambda|}}{\sqrt{|2\pi \Cm|}} \frac{\exp(-\frac{1}{2}\bX^T \Cm^{-1}\bX)}{\exp(-\frac{1}{2}\bY^T \Lambda^{-1}\bY)} \,Z(\bY).
 \]
 
\subsection{Cosmological example --- the Pantheon+SH0ES dataset}
\label{Pantheon}

\begin{figure}
    \centering
\hspace{-0.0cm}
    \includegraphics[width=1.1\textwidth]{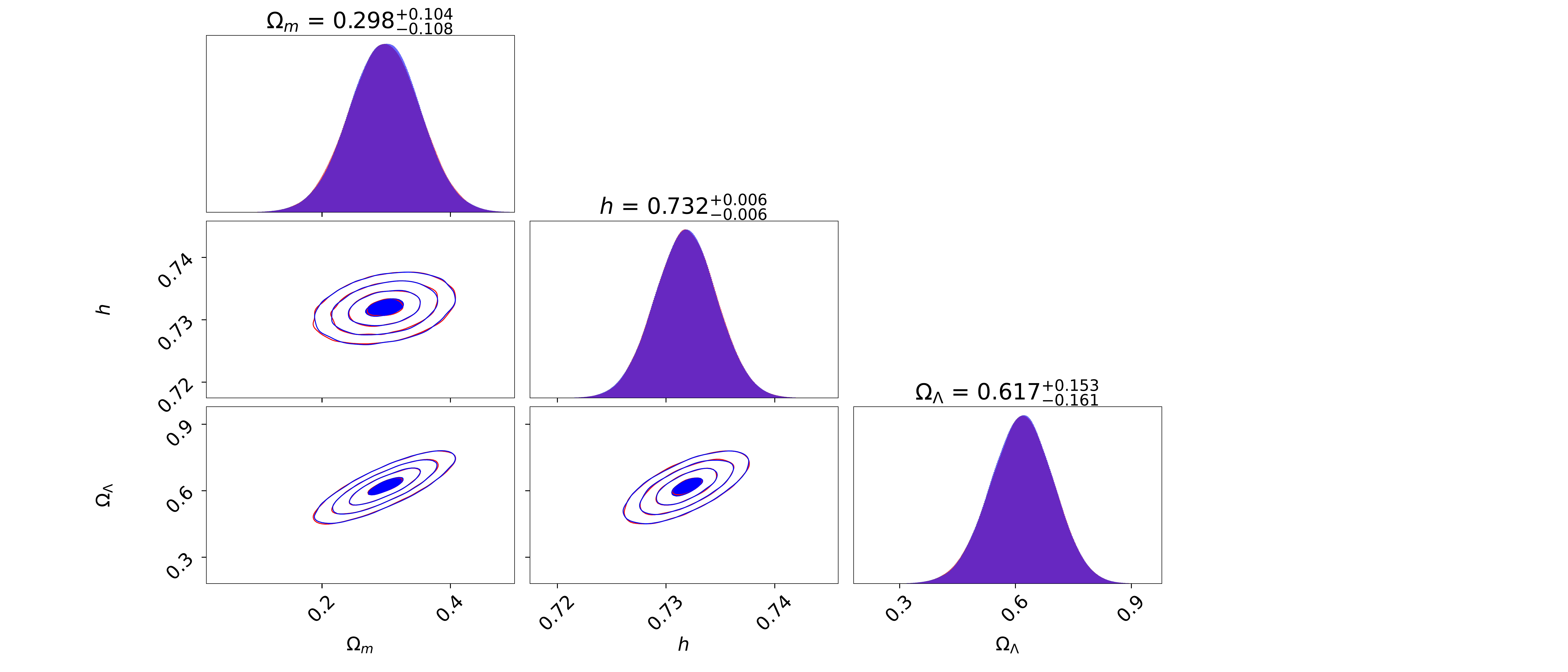}%
    \caption{Marginal posteriors for cosmological parameters $(\Omega_{\rm m},h,\Omega_{\Lambda})$, using the full dataset of size 1590 (red) and only the three MOPED-compressed coefficients (blue). The posteriors are almost identical, resulting in a purple shade in the 1D marginals. }
    \label{fig:SN_post_both}%
\end{figure}

In this section, we re-analyse the Pantheon+SH0ES data presented in \cite{Scolnic2022}.  This consists of 1701 supernovae, with Cepheid calibration.  We use the distance moduli from the catalogue\footnote{https://github.com/PantheonPlusSH0ES} (with a fiducial SNIa magnitude determined from SH0ES 2021 Cepheid host distances), and the supplied covariance matrix, including systematics, to define a simple gaussian likelihood function for 1590 supernovae with $z>0.01$.  While not nearly as sophisticated as a Bayesian hierarchical model \citep[e.g.][]{Shariff2016}, this simplified setup serves to demonstrate parameter inference and Bayesian Evidence comparisons for the full and MOPED-compressed datasets.  The two models being compared are: 
\begin{itemize}
    \item a flat universe, parameterised by $\bbtheta_0 = (\Omega_{\rm m},h)$;
    \item a curved universe, which allows $\Omega_{\Lambda}$ to vary away from $1-\Omega_{\rm m}$, with parameters $\bbtheta_1 = (\Omega_{\rm m},h, \Omega_\Lambda)$,
\end{itemize}
where $h$ is the Hubble-Lemaître constant in units of $100 \, {\rm km}\,{\rm s}^{-1}\,{\rm Mpc}^{-1}$ and $\Omega_{\rm m}$ is the matter density parameter, and $\Omega_{\Lambda}$ the vacuum energy density parameter.
Priors on all varying parameters are taken to be uniform over the range $[0,1]$.  We choose a fiducial model with $\Omega_{\rm m}=0.3$, $h=0.7$ and $\Omega_{\Lambda}=0.7$. The compression concatenates the MOPED coefficients from both models, but since they are nested, this is identical to using the MOPED coefficients from the extended, 3-parameter model, yielding 
\begin{equation}   
{\bf Y}=(1304.68,\ 23857.17, \ -1341.94).
\end{equation}
All the cosmological information about $\Omega_{\rm m}, h$ and $\Omega_{\Lambda}$ is contained in just these three numbers.  The covariance matrix of these coefficients is
\begin{equation}
    {\mathrm {Cov}}({\bY}) = \begin{pmatrix}
        3120.0 & 35028.9 & -2632.9 \\
  35028.9 & 668580.5 & -36364.1 \\
  -2632.9 & -36364.1  & 2559.2
    \end{pmatrix}.
\end{equation}
We use the nested sampling code dynesty\footnote{DOI 10.5281/zenodo.3348367}  \citep{Dynesty2020} with 5000 live points, and no linear approximation is made for the parameter dependence of the mean distance modulus in the likelihood function. We find virtually identical posteriors for full and compressed data, shown in Fig.   
\ref{fig:SN_post_both}. 
The natural log of the Bayesian Evidence for the full dataset is $838.88 \pm  0.035$ for the flat model, and $837.96 \pm  0.037$ for the extended model.  For the MOPED-compressed data, the log Bayesian Evidence values are $-22.35$ and $-23.29$, with the same errors. Hence the Bayes factor are
\begin{eqnarray}
    -0.92 &\pm& 0.05; \qquad \mathrm{Full \ dataset\ (from \ 1590\ numbers)}\nonumber\\
     -0.94 &\pm& 0.05; \qquad \mathrm{MOPED\ compression\ (from \ 3\ numbers),}
     \label{eq:Bayes_MOPED}
\end{eqnarray}
with in both cases the flat model being slightly favoured by the data.  The error is an estimate from the values of about $0.036$ for each, since we do not know the extent to which the errors are correlated.

Fig. \ref{fig:SN_post_both} and the Bayes factors of eq.(\ref{eq:Bayes_MOPED}) show the strength of MOPED for both parameter inference and now for model comparison as well.  The latter, along with the sampling distributions of the Bayes factor in the compressed case (Fig. \ref{fig:BXY_nonlin}) are the main new results of this paper.

\section{Discussion and Conclusions}
\label{sec:conclusions}

In this paper we have shown how the extreme data compression algorithm MOPED \citep{2000MNRAS.317..965H} can be used to compute Bayes factors with essentially no loss of accuracy, thus extending its power from parameter inference to now include model comparison.   The key to this is to perform MOPED compression in all the model spaces, and to concatenate the MOPED coefficients into a slightly larger compressed data space, of dimension equal to the sum of the number of parameters if the models are not nested.  This is still an extreme compression, if the number of data far exceed the number of parameters, which is typically the case with cosmological survey data. In the nested case the MOPED coefficients come from the more complex model. Many of the advantages extend to other forms of score compression when MOPED is not applicable.

The computation of Bayes factors is generally expensive, requiring integrals over the parameter space or sophisticated nested sampling techniques, so where MOPED leads to much faster likelihood evaluations from the massively reduced dataset, the process may be significantly accelerated.  

Finally, we have investigated the sampling distribution of the Bayesian Evidence and of Bayes factors, for uncompressed and MOPED compressed data. We found that while extreme data compression reduces the variance of the Evidence for each individual model, as previously claimed, the sampling distribution for the Bayes factor remains unaffected by extreme compression of linear models, and virtually identical for nonlinear models.  Thus treating the Bayes factor as a frequentist statistic, it is just as effective to compute it from the compressed dataset as the full set.   In a cosmological application,  we have shown that, assuming a simple gaussian data model for the Pantheon+SH0ES dataset \citep{Scolnic2022}, the Bayes factor for the flat vs non-flat model can be computed from only 3 compressed numbers, rather than the full dataset of 1590 supernovae, with no loss of accuracy.

\section*{Acknowledgements}

We are grateful to Andrew Jaffe, David van Dyk, Louis Lyons, and Alex Geringer-Sameth for useful discussions. RT acknowledge co-funding from Next Generation EU, in the context of the National Recovery and Resilience Plan, Investment PE1 – Project FAIR ``Future Artificial Intelligence Research''. This resource was co-financed by the Next Generation EU [DM 1555 del 11.10.22]. RT is partially supported by the Fondazione ICSC, Spoke 3 ``Astrophysics and Cosmos Observations'', Piano Nazionale di Ripresa e Resilienza Project ID CN00000013 ``Italian Research Center on High-Performance Computing, Big Data and Quantum Computing'' funded by MUR Missione 4 Componente 2 Investimento 1.4: Potenziamento strutture di ricerca e creazione di ``campioni nazionali di R\&S (M4C2-19 )'' - Next Generation EU (NGEU). This publication was made possible through the support of an LSSTC Catalyst Fellowship to AM, funded through Grant 62192 from the John Templeton Foundation to LSST Corporation. The opinions expressed in this publication are those of the authors and do not necessarily reflect the views of LSSTC or the John Templeton Foundation.

\section*{Data Availability}
 
No new data are used in this publication, except for those generated randomly by the processes described fully within.



\bibliographystyle{JHEP}
\bibliography{BFrefs} 







\label{lastpage}
\end{document}